\documentclass[%
 reprint,
 amsmath,amssymb,
 aps,
 pra
]{revtex4-2}

\usepackage{graphicx}
\usepackage{dcolumn}
\usepackage{bm}
\usepackage{hyperref}

\usepackage{upgreek}

\begin{document}

\title{Mirror Surface Nanostructuring via Laser Direct Writing - Characterization and Physical Origins}
\author{Mario Vretenar}
\author{Jan Klaers}

\affiliation{
 Adaptive Quantum Optics (AQO), MESA$^+$ Institute for Nanotechnology, University of Twente, PO Box 217, 7500 AE Enschede, Netherlands
}
\date{\today}
             
\begin{abstract}
The addition of an optically absorptive layer to otherwise standard dielectric mirrors enables a set of laser direct writing nanostructuring methods that can add functionality to such mirrors while retaining their high reflectivity. These mirrors are particularly suited for use in optical microcavities, where arbitrary potential landscapes for photons may be constructed. Experiments with photon Bose-Einstein condensates, where high cavity finesse is essential, is one area that has greatly benefited from this approach. A thorough characterization of our implementation of this method is given in this paper, and its physical origins are investigated. In particular, our measurements show that laser direct writing of such mirrors has a reversible and a permanent component, where the reversible process originates from the thermal expansion of the surface and allows a simple yet precise way to temporarily modify the shape of the mirror. Scanning electron microscope cross-sectional images suggest that the permanent part of the nanostructuring process is due to thermally induced pore formation and enlargement in the tantalum oxide layers of the used dielectric mirror.
\end{abstract}

\maketitle
\section{Introduction}\
The ability to manipulate the surface profile of dielectric mirrors while maintaining high reflectivity is particularly useful for creating arbitrary potential landscapes for light in high-finesse optical microcavities \cite{flatten2016}. Suitable methods resulting in permanent potentials include focused ion beam milling \cite{Walker2021,trichet2015}, laser ablation \cite{greuter2014} and controlled buckling \cite{Epp2010, allen2011}. The former two involve substrate milling, followed by dielectric stack coating, while in the latter a polymer is introduced into select regions within the stack, causing local delamination upon thermal annealing. A drawback of these methods is that they require sophisticated cleanroom techniques. This issue is overcome in a recently demonstrated nanostructuring method that uses laser direct writing on dielectric mirrors with an additional absorptive layer to create surface structures with angstrom-level precision \cite{Kurtscheid2020}. This method preserves the high reflectivity of the mirrors, as was previously demonstrated by cavity ring down measurements \cite{Kurtscheid2020}. In this paper, we detail our implementation and advances to this method, perform time-resolved measurements of the process, and analyze its physical origins.\\ 

Thanks to this novel nanostructuring method, open-access optical microcavities become a very flexible platform for experiments with two-dimensional photon gases, in particular, for the study of lasing phenomena and photon Bose-Einstein condensation (BEC) \cite{klaers2010}. The latter may occur when cavity losses are low enough to allow the photons to thermalize to the dye temperature \cite{Schmitt2015}. Experiments with photon BECs in arbitrary potential landscapes realized in this manner provide insights into the condensation process in open quantum systems \cite{Vretenar2021_2}, coupling of photon BECs \cite{kurtscheid2019, Vretenar2021_1, Toebes2022}, and compressibility \cite{, busley2022}. A major future application may be spin glass simulation, allowing the system to be used as a computing platform. Furthermore, polariton systems based on open cavities \cite{scafirimuto2021,feng2022} may benefit from this method. Beyond studying condensation phenomena, nanostructuring of high-reflectivity mirrors via laser direct writing may also find novel applications in areas such as cavity ring-down spectroscopy \cite{berden2000}, wavefront shaping \cite{rotter2017} and others.

In a paraxial approximation, the photons in an optical microcavity follow a modified energy-momentum relation given by
$$E \simeq \frac{mc^2}{n_0^2}+\frac{(\hbar k_r)^2}{2m}-\frac{mc^2}{n_0^2}\left( \frac{\Delta d}{D_0}+\frac{\Delta n}{n_0}\right),$$
where $m$ is an effective photon mass, $k_r$ corresponds to the in-plane wavenumber, and the third term in the equation above corresponds to the potential energy. Here $D_0$ is the cavity spacing, $n_0$ the refractive index of the cavity medium, while $\Delta d(x,y)$ and $\Delta n(x,y)$ represent small variations in the mirror surface height profile and the refractive index, respectively. A local change in the distance between the mirrors or a change in the refractive index thus causes a change in the effective potential energy to which the light is exposed. A common challenge in microcavity systems is the \emph{in-situ} control of the flow of light inside the cavity by controlling this potential energy landscape. Mirrors with integrated piezo arrays \cite{Wlodarczyk2014}, heater arrays \cite{Canuel2012} and electrostatic membranes \cite{bonora2012} may be used for this purpose, however, with limited lateral resolution. As in the case of permanent nanostructuring \cite{Kurtscheid2020}, absorptive layers in the mirror also offer a solution to the problem of temporary changes in the potential landscape. This includes refractive-index tuning via laser writing of the thermo-responsive polymer Poly(N-isopropylacrylamide) (pNIPAM)\cite{Dung2017} added to the cavity medium, and thermal expansion of the mirror surface, which will be demonstrated in this paper. The former introduces an attractive potential, while the latter creates a repulsive potential. Since both techniques work on different timescales, they may be used in a complementary manner. Additionally, a certain degree of spatial complexity in the generated potential landscape can be introduced by wavefront shaping the writing laser.  

The dielectric mirrors used in this work are produced using ion beam sputtering (IBS), and have the following composition: a fused silica substrate, atop which lies an optically absorptive amorphous silicon layer of 30 nm thickness, followed by a quarter-wave dielectric stack with a reflectivity maximum at a wavelength of 600 nm. In most measurements presented in this work, this dielectric stack consists of 20 pairs of Ta\textsubscript{2}O\textsubscript{5} and SiO\textsubscript{2}. However, we also evaluate thinner dielectric stacks with 11 and 10 pairs. The essential point of this mirror design is the absorptive Si layer that provides a way to locally heat the dielectric stack and the optical medium using laser light -- which is the basis of all the phenomena we show in this paper. In the first part of this work, we will  demonstrate that such mirrors enable highly precise methods for permanently and temporarily changing the mirror surface. In the further course of this work, we will explain the physical origins of the method in more detail.

\section{Point-like surface structures}\label{sec-peaks}

\begin{figure*}[tbh!]
\includegraphics[width=\textwidth]{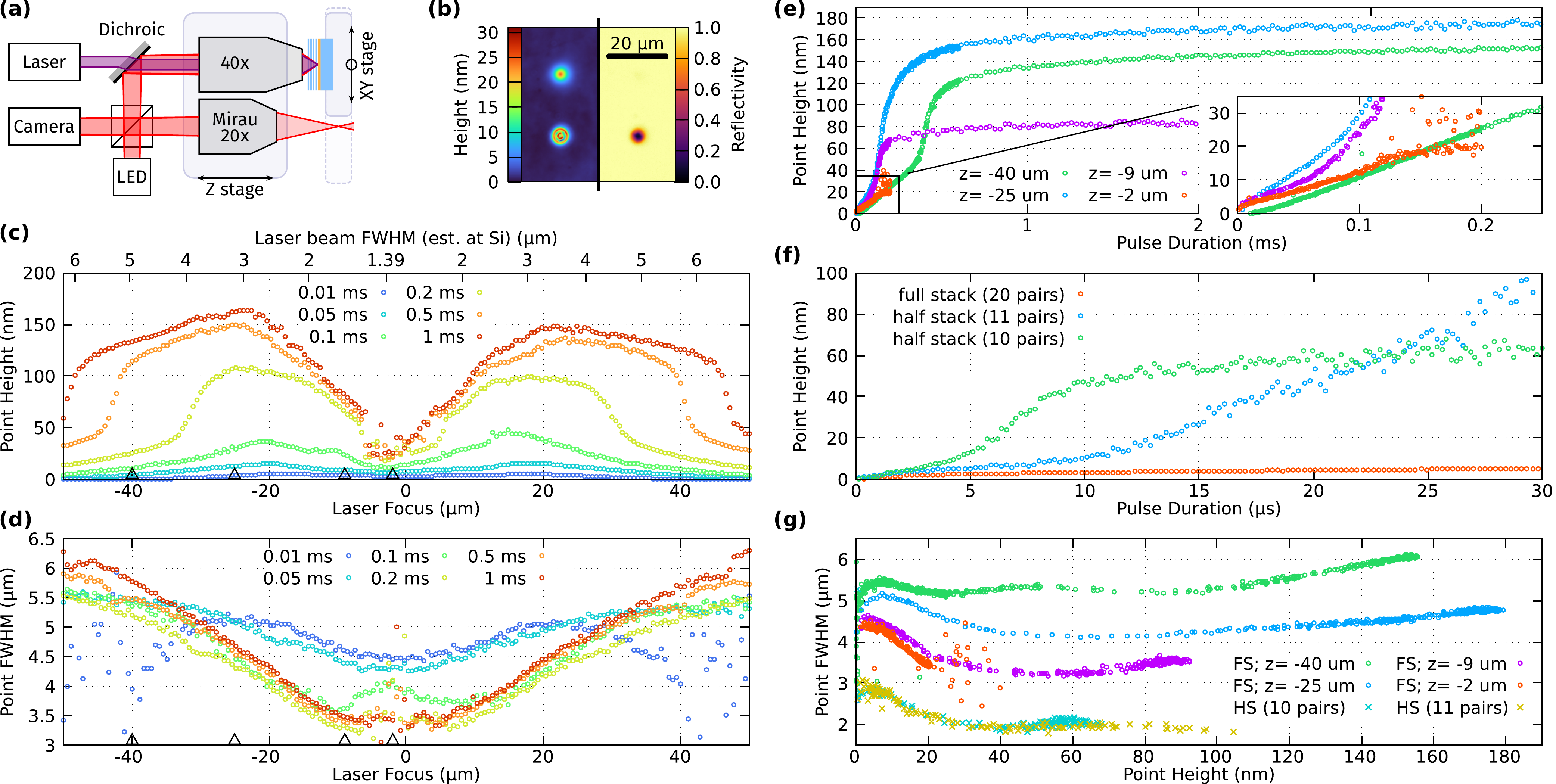}  
\caption{Point-like surface structures. a) Schematic of the nanostructuring setup. b) Height map (left) and local reflectivity (right), of two written point-like structures. The upper structure is a typical example of a written point, and its height profile follows approximately a Gaussian function. The bottom example shows a locally destroyed mirror as a result of excessive irradiance and pulse duration. c) Structure height and d) lateral size (FWHM) as a function of laser focus position relative to the mirror surface for 6 pulse durations. Four different focus positions are marked with triangles. e) Structure height as a function of pulse duration for the four foci, with an inset highlighting the low height (duration) region. f) Structure height as a function of pulse duration for three dielectric stack configurations, with laser focus set at the Si layer. g) Lateral size (FWHM) as a function of structure height for the four foci for the full stack (FS) case, and for the two half stack (HS) mirrors (in-focus writing only). In graphs c-g) each data point corresponds to one written structure.}
\label{fig:points}
\end{figure*}

A schematic of the nanostructuring setup is shown in \textbf{Figure \ref{fig:points}}a. It consists of two main parts: a white light Mirau interferometer used for obtaining the mirror surface height profiles, and the direct laser writing part. A \mbox{200 mW} continuous wave (CW) laser operating at 405 nm is used as the writing laser, impinging on the sample mirror from the front side with an estimated beam waist of 1.39 \textmu m. An initial investigation of the nanostructuring procedure is performed by writing a series of point-like structures while varying focus position and pulse duration. The writing order is somewhat randomized to prevent systematic errors. For each set of points, interferometric surface scans were performed both before and after the writing, with the analysis performed on the height difference in order to minimize the effect of the residual mirror surface roughness. \textbf{Figure \ref{fig:points}}b shows the height map and reflected light intensity in two distinct cases: at the top, a point-like surface structure of approximately 25$\,$nm  height, and at the bottom an example of a locally destroyed mirror, which may occur for sufficiently high irradiance. The height profiles of the point-like structures are generally observed to follow a Gaussian distribution with a certain lateral extent.

First, we examine the dependence of the structure height and lateral size on the laser beam focus position relative to the mirror surface, see \textbf{Figures \ref{fig:points}}c and \textbf{\ref{fig:points}}d, respectively. Six characteristic pulse durations are used. Both figures show a symmetry axis around a focus position of $z\approx -3\,$\textmu m, which is close to the location of the Si layer. Varying the focus position affects the writing process by altering the beam width at the position of the thin silicon layer (shown as a second x-axis in \textbf{Figure \ref{fig:points}}c), and thus the irradiance. Writing in focus with sufficiently long pulses locally destroys the mirror, which results in missing or scattered data points. The smallest achievable lateral structure size is 3.5 \textmu m (FWHM), which is roughly three times larger than the laser beam focus size. Overall, the lateral size of the written structures is somewhat independent of the diameter of the writing beam at the silicon layer. In particular, the minimum achievable feature size does not seem to be limited by the beam diameter. Four focus positions are chosen for further calibration: $z=$ -2, -9, -25 and -40 \textmu m, in the following text they are referred to as 'calibration foci', as marked by triangles on the focus position axis.

The dependence of the structure height on the pulse duration for the four calibration foci is shown in \textbf{Figure \ref{fig:points}}e. For near-focus writing ($z=-2\,$\textmu m), the curve is incomplete because for the given laser power, pulse durations longer than 0.2 ms have a chance of causing local destruction to the mirror, as seen in \textbf{Figure \ref{fig:points}}b bottom. Such events eject material from the mirror which lands in the surrounding area, contaminating other structures. The spread of observed point heights just before this phenomenon starts to occur indicate the onset of mirror destruction. The acquired data using the three other foci show that the structure height asymptotically tends toward a maximum value specific for that particular focus. The maximum achievable height is almost 200 nm, when writing with a focus position of about \mbox{20 \textmu m}. The observed trend that the structure height increases with the beam size of the writing laser at the silicon layer suggests that even higher structure heights might be achieved for higher laser power. 

The role of the thickness of the quarter-wave stack on the structure height at given pulse durations is investigated in \textbf{Figure \ref{fig:points}}f. In this case, all writing procedures are done in-focus to the silicon layer. The first striking difference is that the pulse durations required to achieve certain heights for thinner dielectric stacks are an order of magnitude lower than those for the thicker stack. Furthermore, a significant reduction is observed even with the removal of just one extra layer pair. In the case of 11 pairs, a pulse duration just above \mbox{30 \textmu s} starts to destroy the mirror. The lateral size (FWHM) of the written structures as a function of their height is shown in  \textbf{Figure \ref{fig:points}}g.  In general, lower structures are also wider with the FWHM being about \mbox{4.5 \textmu m} for in-focus writing. As the structure height increases, the FWHM falls down to a value of about \mbox{3.3 \textmu m}, after which it rises again. Interestingly, the FWHM of written structures on a 10(11)-pairs quarter-wave stack is almost half compared to in-focus nanostructuring of the 20-pairs stack. Significantly smaller lateral structural sizes can thus be implemented in this way for applications that do not place excessively high demands on the reflectivity.

\section{Plateau-like surface structures}\label{sec-plateaus}

\begin{figure}[tbh!]
\includegraphics[width=\columnwidth]{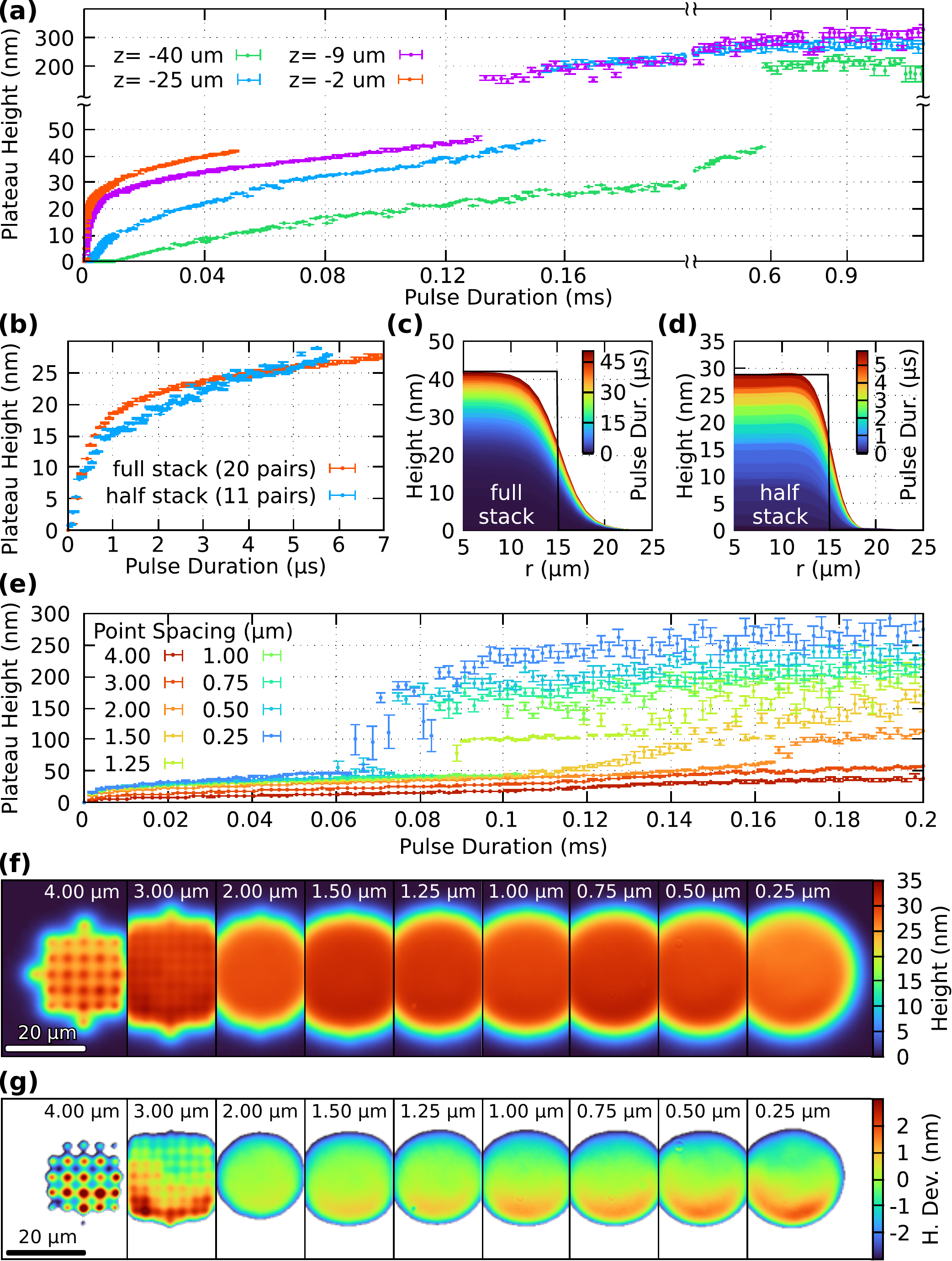}  
\caption{Plateau-like structures. a) Plateau height as a function of pulse duration for the four calibration foci in the case of a 20-pairs stack and b) for the 11-pairs stack (near-focus writing only). Each data point corresponds to one written plateau-like structure. The determined heights are an average of the height profile within half the plateau radius, with the error bars representing their standard deviation. c) The radial height profile for near-focus writing in case of the 20-pair stack and d) the 11-pair stack, with colors representing different pulse durations. The black lines show the intended plateau circle radius. e) Plateau height as a function of pulse duration for 9 point spacings using a -9 \textmu m focus position. f) Height maps of plateaus of about 30 nm height for different point spacings. g) Replot of the data from f) highlighting the height deviation from the plateau average.}
\label{fig:plateaus}
\end{figure}

As complex structures involve writing a grid of overlapping points, writing plateaus of various heights is a reasonable starting point towards more arbitrary surface structures. The writing and interferometric surface scan procedure is the same as in the previous section, except for the fact that now the point spacing/density shows up as an additional parameter. Circular plateaus are written at the four calibration foci with various pulse durations, and a fixed point spacing of \mbox{1 \textmu m} is used unless noted otherwise. The writing is done from the bottom corner up.

The resulting plateau height as a function of pulse duration is shown in \textbf{Figure \ref{fig:plateaus}}a. Low pulse durations result in a behavior similar to that observed for point-like structures, however, only up to a maximum height of about 45 nm for all tested focus positions. Above that, in the case of near-focus writing we observe mirror destruction, while for other focus positions a sharp jump occurs suggesting the onset of delamination. The delaminated portion has a bubble-like shape, with the total height appearing to be limited by the plateau radius. We furthermore write plateaus using the 11-pairs stack, with the results shown in \textbf{Figure \ref{fig:plateaus}}b. The maximum achievable height is about \mbox{25 nm} which is lower compared to the previous value of \mbox{45 nm}.
In contrast to point-like structures, the plateau height as a function of pulse duration is observed to be less dependent on the size of the dielectric stack.
The radial height profiles are determined and plotted as a function of pulse duration in \textbf{Figures \ref{fig:plateaus}}c and \textbf{\ref{fig:plateaus}}d for the 20-pairs and 11-pairs dielectric stack, respectively. The profiles show a blurring of the boundaries of the plateau, which is more pronounced for the thicker dielectric stack. In the case of out-of-focus writing, similar but wider profiles are obtained. 

Next, we evaluate the effect of the point spacing on the nanostructuring of extended structures. Circular plateaus were written for 9 different point spacings, using a $z=-9\,$\textmu m focus position. The plateau height as a function of pulse duration is shown in \textbf{Figure \ref{fig:plateaus}}e. The previously determined maximum height of around 45 nm before the onset of delamination is observed to apply for all investigated point spacings. The height standard deviation within each plateau, as marked with the error bars, indicates how uniform the surfaces are. The height maps of 30$\,$nm high plateaus written with different point spacings are shown in \textbf{Figure \ref{fig:plateaus}}f revealing that, for this particular focus position, point spacings equal or smaller than 2 \textmu m already provide relatively smooth structures. 
The height deviation from the averaged plateau height is shown in \textbf{Figure \ref{fig:plateaus}}c, revealing that the plateaus start to become increasingly inhomogeneous with denser point spacing. This suggests that in the case of spatially overlapping writing processes, the relationship between structure height and pulse duration changes (the writing is done bottom to top). This effect is less noticeable with lower \mbox{($\lesssim20$ nm)} structures.\\

\section{Complex structures}\label{sec-complex}

\begin{figure}[tbh!]
\includegraphics[width=\columnwidth]{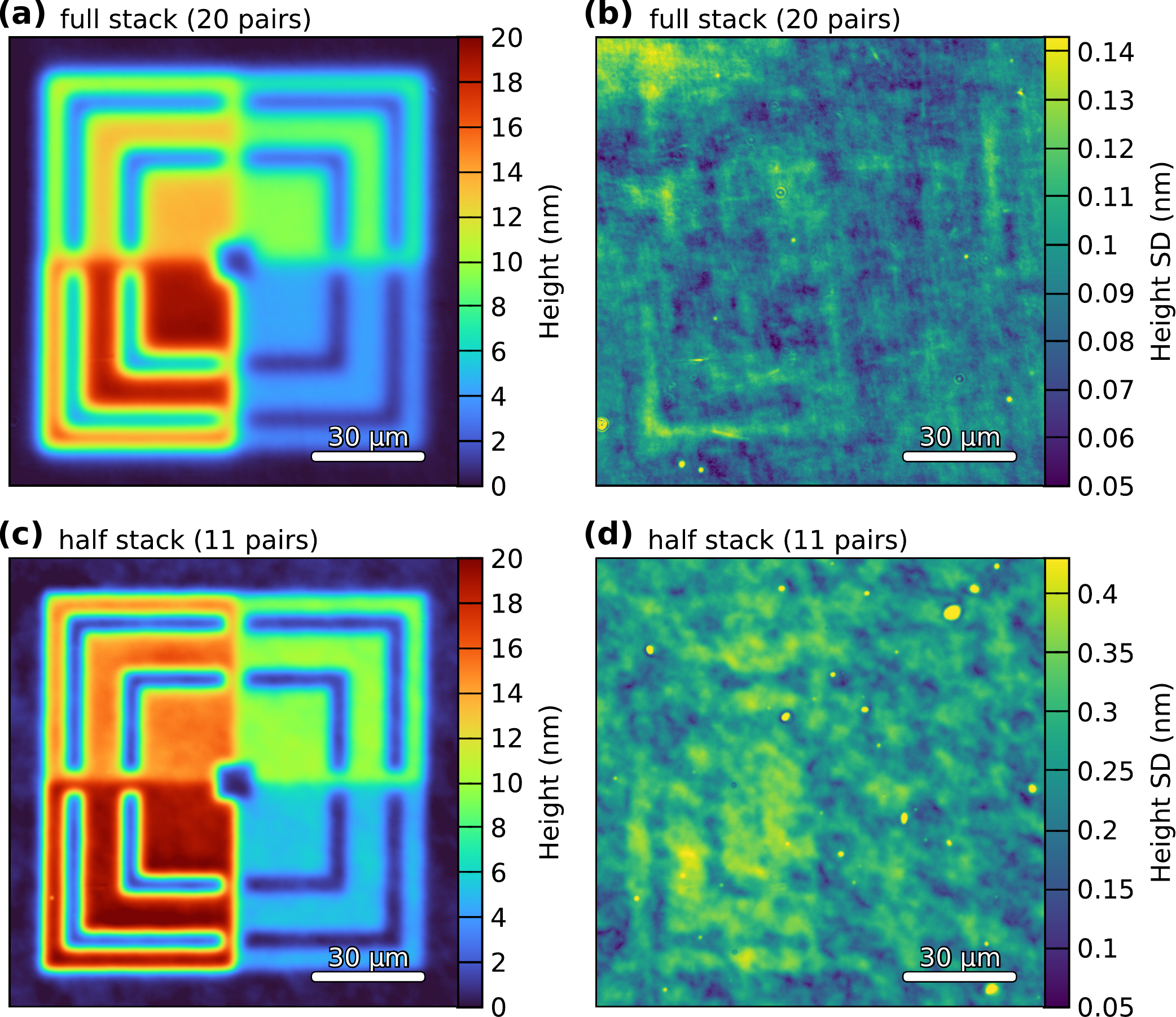}  
\caption{Height resolution of the nanostructuring process. a) Measured height map of the written test structure and b) the height standard deviation of 10 written test structures, written on the full (20-pairs) stack. c) Measured height map of the written test structure and d) the height standard deviation of 5 written test structures, written on the half (11-pairs) stack.}
\label{fig:complex}
\end{figure}

In order to write arbitrary structures, the data shown in \textbf{Figures \ref{fig:plateaus}}a and \textbf{\ref{fig:plateaus}}b is used to define a mapping between structure height and pulse duration (by fitting B-splines to the experimental data). This mapping can then be used to convert a target height map to a pulse duration map that can be written with the nanostructuring setup. 
For the purpose of testing the capabilities and repeatability of the so-defined nanostructuring procedure, we design a simple test structure featuring four height levels of 5, 10, 15 and 20 nm with sharp features of various lateral sizes. Writing the structure using the \mbox{-2 \textmu m} focus calibration results in the height profile shown in \textbf{Figure \ref{fig:complex}}a. Here, unlike in the previous sections, no background subtraction was made and correspondingly the height map also contains the mirror roughness. The obtained height map is an average of 10 interferometric scans. Some of the narrow features appear blurred or do not reach the intended height due to their small lateral size (compared to the point spread function of the writing process). The repeatability of the nanostructuring is tested by writing 10 identical structures, and calculating the standard deviation in the obtained height. The result is shown in \textbf{Figure \ref{fig:complex}}b, indicating a standard deviation below 0.1 nm which matches the surface roughness of the mirror.
These results show that the nanostructuring process is precise down to the level of single atomic layers. Test structures were written with the other three focus positions, and they show some additional feature broadening as expected. The repeatability, however, is the same. We furthermore repeat the writing process on the thinner 11-pairs dielectric stack, with the resulting height profile shown in \textbf{Figure \ref{fig:complex}}c. Here the features are noticeably sharper: the thin edge stripes reach the intended height and the valleys seem to reach the expected baseline as well. Five identical structures were written, and the standard deviation is shown in \textbf{Figure \ref{fig:complex}}d. In this case, a mirror with a rougher substrate was used, and the measured deviations again match the overall surface roughness. 

\section{Long-term stability}\label{sec-lts}

\begin{figure}[tbh!]
\includegraphics[width=\columnwidth]{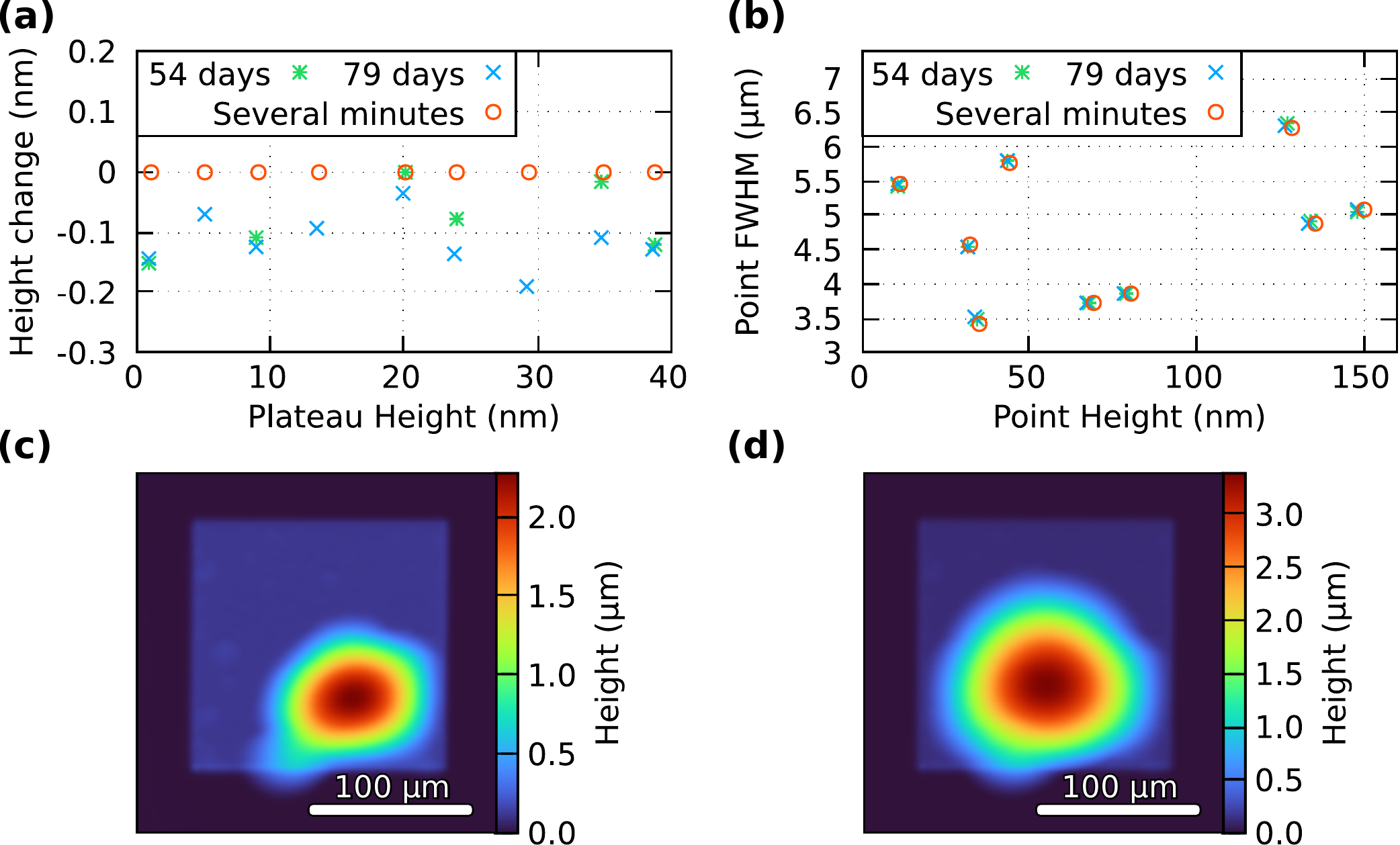}
\caption{Long-term stability of surface structures. a) Observed height change in 54 and 79 days compared to a few minutes after nanostructuring for plateau-like structures written at a -2 \textmu m focus position. Note: some of the heights were not evaluated at the 54 day mark. b) FWHM and height of point-like structures for three focus positions (-9, -25 and -40 \textmu m) and three pulse durations (0.2, 1 and 5 ms). c) Height map of a structure realized by writing an uniform square showing delamination, and d) height map of the same structure taken 11 days later.}
\label{fig:stability}
\end{figure}

Surface measurements performed at intervals of a few months show that the nanostructuring is fairly stable with a slight shrinkage in the order of the measurement error, as can be seen in \textbf{Figure \ref{fig:stability}}a for plateau-like structures. This holds as long as the nanostructuring has been carried out in the non-delaminating regime. Point-like structures of various focus positions and durations also show good long-term stability both in height and lateral size (FWHM), see \textbf{Figure \ref{fig:stability}}b, with larger deviations present beyond the delamination threshold. In the delamination regime, on the other hand, our measurements show that the written structures exhibit a significant change over a period of a few days. Writing an uniform square using the -40 \textmu m focus position with long pulse durations resulted in the structure shown in \textbf{Figure \ref{fig:stability}}c, which features a bubble-like structure of 2.5 \textmu m height near the bottom-right corner. The same structure was scanned again 11 days later, with the height map shown in \textbf{Figure \ref{fig:stability}}d. The observed growth of the bubble in both height and lateral size indicates the presence of a slow relaxation timescale in the delamination process of laterally extended structures.

\section{Time-resolved measurements} \label{sec-trm}

\begin{figure*}[tbh!]
\includegraphics[width=\textwidth]{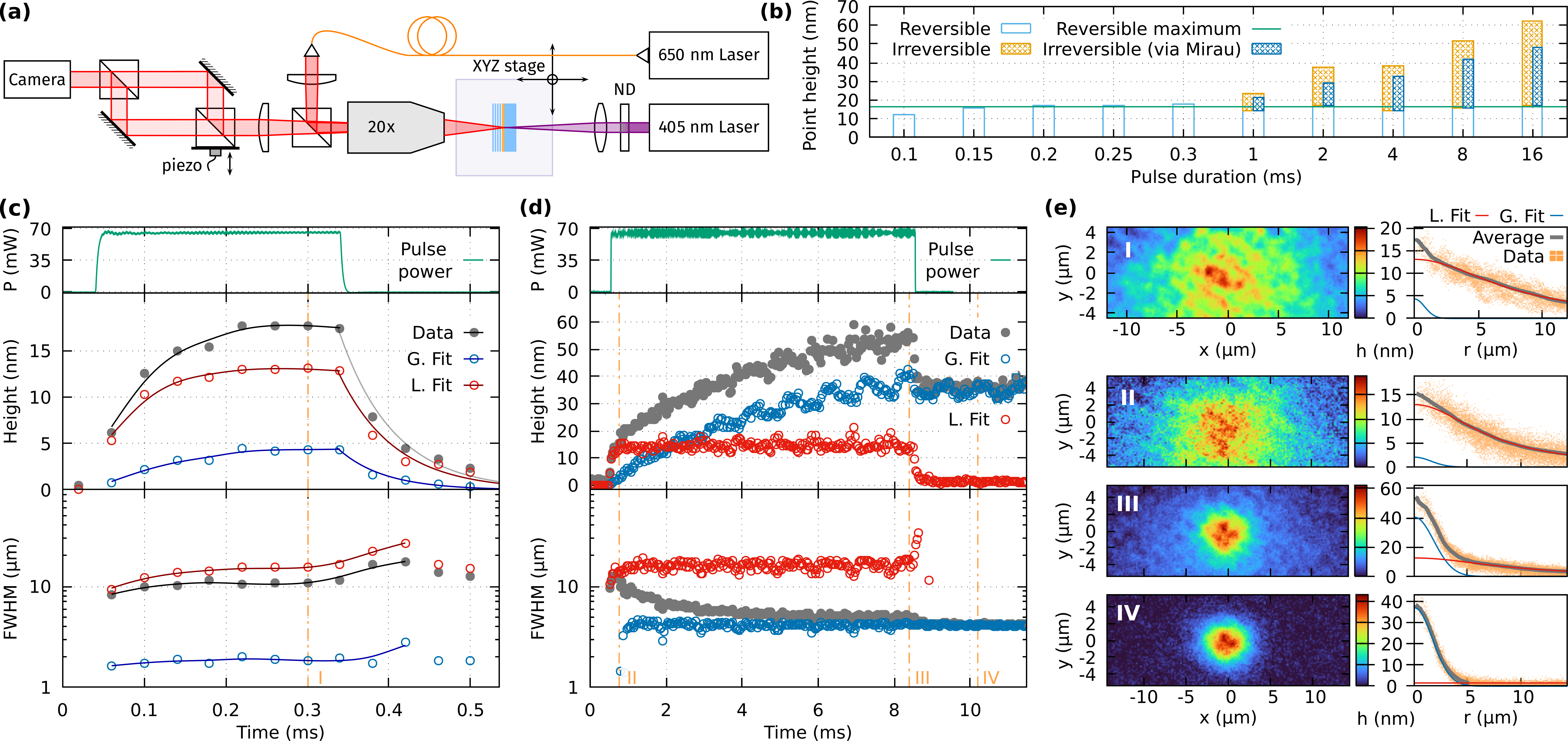}  
\caption{Time-resolved measurements. a) Schematic of the time-resolved measurement setup. b) Overview of the time-resolved measurement results showing the maximum achieved height during the nanostructuring process, as well as the permanent components. c) Results for a pulse duration of 0.3 ms, where no permanent nanostructuring occurs, showing the pulse power, measured structure height, and lateral size (FWHM) as a function of time. d) Results for a pulse duration of 8 ms, where permanent nanostructuring is achieved. e) Typical height profiles in cartesian (left) and radial (right) coordinates at various times during the writing process, as marked in c) and d).}
\label{fig:timeres}
\end{figure*}

To better understand the nanostructuring process, we perform time-resolved measurements of the surface profile during and after the heating pulses.
A simplified schematic of the experimental setup is shown in \textbf{Figure \ref{fig:timeres}}a. A \mbox{405 nm} laser is used as the laser source with 65 mW of optical power focused (from the back side) down to an estimated beam waist of \mbox{$7$ \textmu m} on the mirror's silicon layer. A portion of that light is captured by a photodiode, providing the temporal measurement of the pulse power. A Mach-Zehnder interferometer (MZI) was constructed with the purpose of interfering the image of the surface with itself, offset vertically by a set amount. In this way, a flat region of the sample mirror is used as a reference plane for the interferometric height determination. A camera captures snapshots of a narrow horizontal strip with an effective temporal resolution of 40 \textmu s, which is reached by combining multiple measurements in a stroboscopic fashion. Height maps are then calculated for each image, after a certain degree of noise removal. 

Ten pulse durations were tested: 0.1, 0.15, 0.2, 0.25, 0.3, 1, 2, 4, 8 and 16 ms, with the pulse power over time closely following a square profile.  The last five durations result in permanent structures. Height and lateral size (FWHM) of the structure for 0.3 ms pulse duration is shown in \textbf{Figure \ref{fig:timeres}}c, which is the longest pulse duration that does not lead to a permanent structure. 
The data points are connected using splines as a guide to the eye, except in the tail where an exponential fit is used. Despite the absence of a permanent structuring, it can be observed that heating the mirror with the focused laser beam creates a temporary surface elevation that can be as high as 17 nm. The creation and relaxation of this elevation occurs on the millisecond timescale, which is consistent with a thermal expansion process. The lateral feature size is observed to be more than twice than that of the permanent structures.
Further analysis of the measured height profiles, shown in \textbf{Figure \ref{fig:timeres}}e, indicates that the observed profiles can be modeled by superimposing a broad Lorentzian background ('L. Fit') with a narrower Gaussian profile ('G. Fit'). These two contributions are resolved in \textbf{Figure \ref{fig:timeres}}c as a function of time.
The usefulness of this data modeling is not obvious at first, but becomes apparent in the case of permanent nanostructuring with longer pulse durations.
The measurement for 8 ms pulse duration, for example, is shown in \textbf{Figure \ref{fig:timeres}}d. With this pulse duration, a permanent structure of about 35 nm height is created. Interestingly, it is observed that this permanent structure corresponds exclusively to the narrower Gaussian component of the height profile. The broader Lorentz-like component profile quickly disappears after the writing pulse power falls to zero. This circumstance corresponds to the aforementioned observation that the lateral structure size in the case of a reversible (thermal) expansion of the mirror surface is significantly larger than that of the permanent structuring.

The maximum height obtained within one pulse for the ten pulse durations that were tested is shown in \textbf{Figure \ref{fig:timeres}}b, with the reversible and the permanent components indicated separately. The reversible component consistently saturates around 17 nm. The blue hatched bars show the structure heights as determined by the Mirau interferometer in the setup shown in \textbf{Figure \ref{fig:points}}a. The observed mismatch is most likely due to an actual relaxation of the written structures shortly after the writing process.

\section{Physical origin of the nanostructuring process} \label{sec-pnp}

In this section, we present experimental evidence that the permanent part of the nanostructuring method relies on thermally induced pore formation and enlargement in the tantalum oxide layers of the used dielectric mirrors. 
The amorphous oxide films used in the quarter-wave stack of our mirrors, Ta\textsubscript{2}O\textsubscript{5} and SiO\textsubscript{2}, are produced via IBS using argon ions. These amorphous thin films are widely used and have been studied frequently \cite{rubio1982,Cevro1995,chaneliere1998,bartic2002,penn2003,zhao2003,Paolone2022}. In particular, it is known that amorphous tantalum oxide layers produced by IBS store a high degree of mechanical energy, which can be released by annealing processes, whereby the layer thickness of the films increases by a few percent \cite{anghinolfi2013, Paolone2021}. Seen microscopically, this growth comes about through an enlargement of pore structures within the amorphous film. For SiO\textsubscript{2}, this effect also present but is significantly smaller \cite{granata2020amorphous}.
While the mentioned results involve long annealing times, higher temperatures as they occur in the here investigated writing process may produce a similar effect on much shorter timescales.\\

\begin{figure}[tbh!]
\includegraphics[width=0.98\columnwidth]{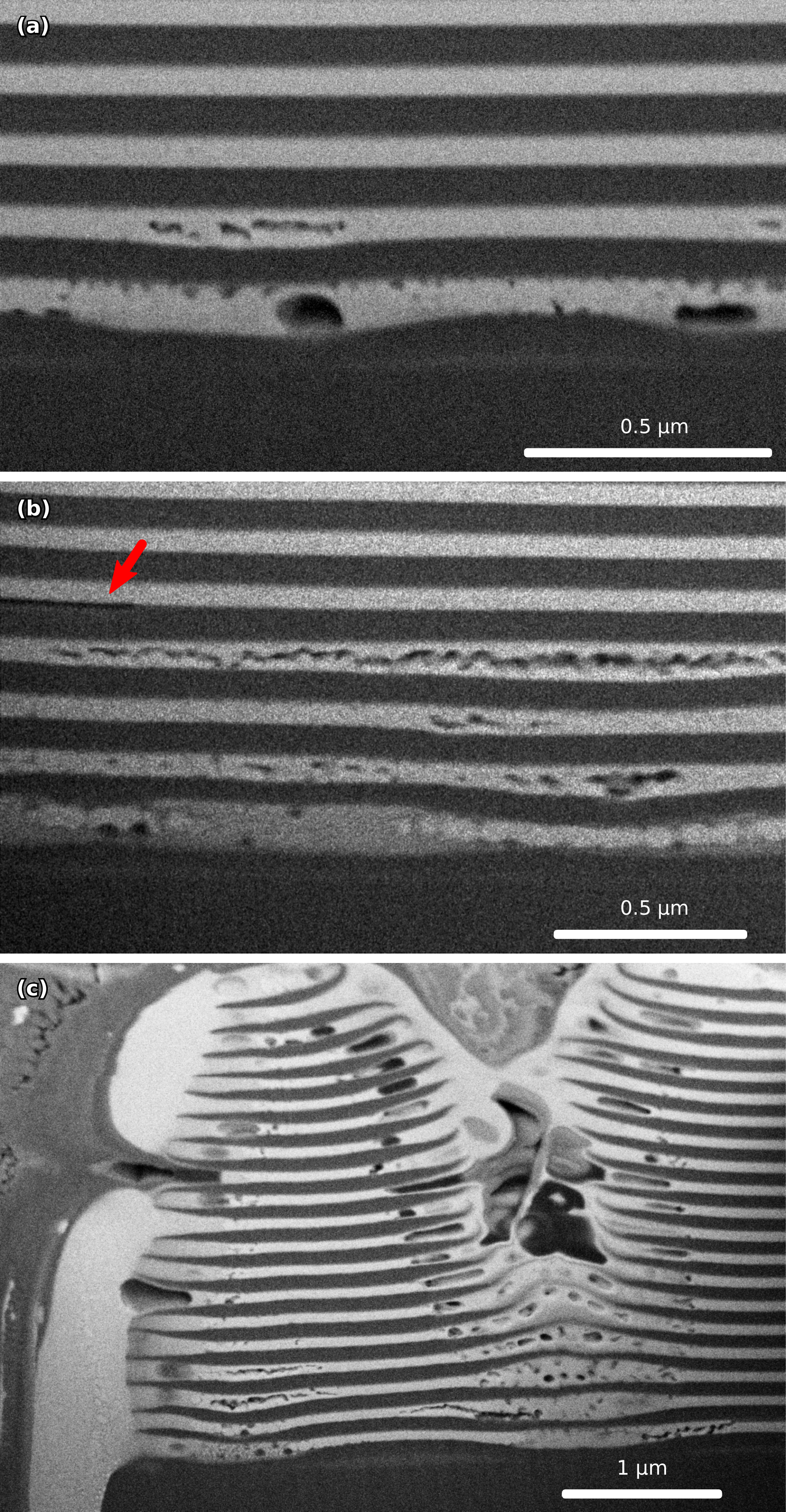}  
\caption{SEM images of written surface structures. The mirror cross sections created via FIB milling are shown for a) a 40 nm high plateau written at $z=-2\,$\textmu m, and b) a delaminated plateau written at $z=\,-9$\textmu m, both using a \mbox{1 \textmu m} point spacing. The lowermost layer is the substrate. Bubbles in the 10-100 nm size range can be clearly seen within the first few $\text{Ta}_2\text{O}_5$ layers, but not within $\text{SiO}_2$. In the delaminated case in b), a smooth crack is indicated with an arrow. It extends towards the highly delaminated center of the structure (not shown). c) A cross section of locally destroyed mirror resulting from excessive pulse irradiance. In a-c) the imaging angle is 52\textdegree\ in relation to the mirror normal.}
\label{fig:semfib}
\end{figure}

In order to confirm this hypothesis, we prepare cross sections of several plateau-like structures using focused ion beam (FIB) milling, and image them with scanning electron microscopy (SEM). First, we investigate the non-delaminating regime of the writing process. More specifically, we evaluate a 40 nm high plateau written at $z=-2\,$\textmu m focus position, which is shown in \textbf{Figure \ref{fig:semfib}}a. The SEM image is focused onto the region around the silicon layer. From bottom to top we see the SiO\textsubscript{2} substrate (dark gray), the thin Si layer (mid gray), and then interchanging of SiO\textsubscript{2} (dark gray) and Ta\textsubscript{2}O\textsubscript{5} (light gray) layers. Indeed, pores with sizes in the 10-100 nm range are visible in the bottom-most two Ta\textsubscript{2}O\textsubscript{5} layers, but not in the SiO\textsubscript{2}.
This is consistent with the previously reported results on annealing processes in the thin films concerned, despite the time scale and temperature of the annealing process being somewhat different.
The fact that the nanostructuring takes place in the bottom layers of the dielectric stack explains why the reflectivity of the mirrors is not significantly reduced by the process \cite{Kurtscheid2020}. This also shows the importance of the embedded absorptive Si layer, which guarantees that the heating process takes place at the correct position within the mirror.

Additional SEM images (not shown) of 40 nm high structures written at a -40 \textmu m focus position reveal, however, no visible pores. The images of the cross sections are visually indistinguishable from a non-structured mirror. Due to the fact that the oxide layers are non-conductive, charges build up during SEM imaging which severely limits the resolution, such that feature sizes of \mbox{$\lesssim$ 5 nm} cannot be easily resolved. Furthermore the charge build-up distorts the imaging which makes it difficult to perform a layer thickness analysis \cite{echlin2009}. As previously shown, using the -40 \textmu m focus setting requires an order of magnitude longer pulse durations to reach the same height as compared to the -2 \textmu m focus setting. This longer annealing at lower irradiance appears to result in smaller pore structures that cannot be resolved by SEM, but that lead to the same overall volume increase. Thus, these results do not necessarily contradict the pore hypothesis.

\textbf{Figure \ref{fig:semfib}}b features the cross section of a delaminated plateau structure written at $z=-9\,$\textmu m focus position. 
Two types of delamination can be seen: a long jagged fracture consisting of connected bubbles, and a clean separation of the Ta\textsubscript{2}O\textsubscript{5} and SiO\textsubscript{2} layer as marked with a red arrow. The latter extends towards the center of the plateau structure (i.e. the more strongly delaminated region). A cross section of a crater resulting from writing at the -2 \textmu m focus position and excessive irradiance is given in \textbf{Figure \ref{fig:semfib}}c, showing regions that have melted, bubble formation and cracks.\\

\begin{figure}[tbh!]
\includegraphics[width=\columnwidth]{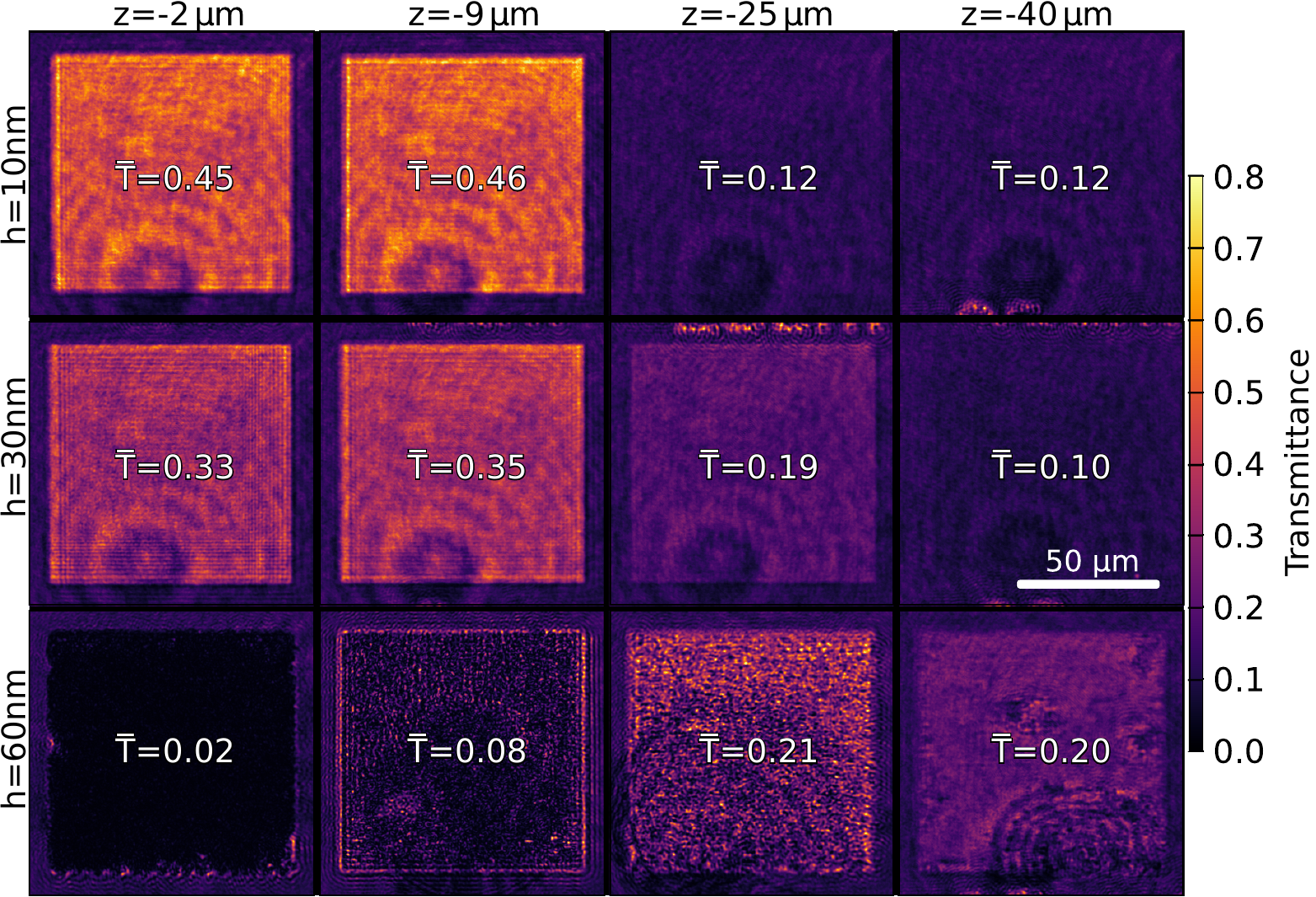}  
\caption{Optical transmittance through nanostructured regions of the mirror. Twelve square plateaus are written using the four calibrations foci (columns) with 3 target heights 10, 30 and 60 nm (rows). The average transmittance of 405 nm light (outside the high-reflectivity region of the mirror) is noted in each structure.}
\label{fig:transm}
\end{figure}

An interesting effect, which is not necessarily related to layer growth but still provides useful information about the physical processes during the writing process, is found in the behavior of the mirror transmission. We observe that the mirror transmittance is altered by the writing procedure for certain parameters. To further investigate this, we perform optical transmission measurements with 405 nm light. For a (blank) mirror with amorphous Si layer, the observed transmittance is 0.10\textpm 0.01. A simple setup consisting of a camera and a microscopic objective is used to measure the transmission of various plateau-like surface structures, with the results shown in \textbf{Figure \ref{fig:transm}}. Structures with three target heights were written using the four calibration focus positions. The transmittance was averaged over the area of each structure, and the resulting values are shown next to the corresponding spatially resolved transmittance maps in \textbf{Figure \ref{fig:transm}}. For reference, transfer matrix calculations provide a transmittance value of 0.07 assuming an amorphous silicon layer in the mirror, and 0.45 for crystalline silicon. We can clearly see that, for structures written with sufficiently high irradiance, the transmittance is closer to that of what is expected for crystalline silicon, than for amorphous. This is consistent with other work, where it has been shown that laser direct writing can cause silicon to crystallize when using CW and nanosecond-scale pulses \cite{Ivanda1991,michaud2006,Cinar2020}.

\section{Conclusion}
In this work, we performed extensive characterization of a novel mirror surface nanostructuring process based on the laser direct writing of dielectric mirrors containing an additional thin absorptive layer \cite{Kurtscheid2020}. The method is characterized by its simplicity, cost-effectiveness and its exceptional precision. Angstrom-like precision in the growth direction can be routinely achieved without expensive clean-room technology. The nanostructuring process is fast, making it an excellent choice for prototyping. The use of this technique has contributed significantly to recent advances in the study of microcavity systems and we believe that other areas of optics can also benefit. This applies in particular to areas in which mirrors with highest reflectivity are used, e.g. cavity ringdown spectroscopy or interferometric precision measurements. Last but not least, wavefront shaping, specifically in the UV, can be a possible application.

\section{Experimental section}\label{sec-exp}
The dielectric mirrors used in this work are manufactured by Laseroptik GmbH using IBS with argon. The amorphous silicon layer added between the substrate and the dielectric stack is typically 30 nm thick. The dielectric stack consists of 20 pairs of Ta\textsubscript{2}O\textsubscript{5} and SiO\textsubscript{2}. With the use of glue we mechanically remove layers from the dielectric stack allowing us to evaluate also the nanostructuring in mirrors with fewer dielectric stack layer pairs, in this case 11 and 10.\\

The nanostructuring setup (see \textbf{Figure \ref{fig:points}}a) consists of two main parts: the white light interferometry and the direct laser writing. Both share the imaging light source - an LED (Thorlabs M617D3) with an emission spectrum situated within the sample mirror's high reflective region, and the camera (Basler ACA1300-200um). The sample mirror is mounted on an XY assembly consisting of two piezo stepping stages with a 13 and 26 mm travel range (Pi N-565.260 and Pi N-565.360). A 13 mm travel range piezo stepping stage (Pi N-565.260) provides the Z axis motion and holds two microscope objectives: a 20x Mirau interferometric objective (Nikon MUL40201) for surface profiling, and a 40x objective (Olympus RMS40X) for the direct laser writing. The foci of the two objectives are intentionally mismatched in such a way that during operation, only one of them has its focus on any surface at any given time. A dichroic is used to insert the writing laser beam from a 200 mW CW laser operating at 405 nm (CNI MDL-NS-405), which is controlled via TTL switching.\\

The time-resolved measurement setup (see \textbf{Figure \ref{fig:timeres}}a) uses a \mbox{405 nm} laser (NEJE B25425) as the heating laser, providing about \mbox{260 mW} of optical power, reduced to 65 mW by a neutral density (ND) filter (OD=0.6). The sample mirror was mounted on a manual XYZ stage with one of the micrometer screws (an axis perpendicular to the laser beam) motorized by a servo. A lens then focuses the light onto the sample mirror's silicon layer. On the imaging side, a standard microscope objective was used instead of the more convenient Mirau interferometric objective in order to avoid possibly damaging the internal reference mirror, as the sample mirror is partially transmissive at \mbox{405 nm}. Instead, a Mach-Zehnder interferometer (MZI) was constructed with the purpose of interfering the image of the surface with itself, offset vertically by a set amount. In this way a flat portion of the sample mirror is used as a reference plane. The first beamsplitter of the MZI is actuated by a piezo in order to control the phase shift. A \mbox{650 nm} fiber tester laser was used as the imaging light source, due to its intermediate coherence length. A narrow horizontal strip of the light exiting the MZI is captured by a camera (Basler acA4024-29um), which allows the camera frame rate to be increased to 1000 frames per second (fps), providing a temporal resolution of 1 ms. Higher temporal resolution is needed, however, which is achieved using a strobe effect where the camera triggers are all offset by a set amount for each individual pulse, and multiple pulses are combined for the final result. With this, a frame rate of 25000 fps, or 40 \textmu s per frame is achievable. In measurements where pulse duration was short enough such that no permanent nanostructuring occurs, 1000 pulses were averaged to produce the final result. With longer pulse durations where permanent structures were created, only 25 pulses per measurement set were performed, and the servo was used to automatically move to a clear portion of the sample mirror after each write.
\\

\section*{Acknowledgements}
M.V. performed the experimental studies and analysis. J.K. supervised the work.\\
We thank Melissa Goodwin for performing the SEM/FIB scans, and Chris Toebes for useful discussions.\\
This work has received funding from the European Research Council (ERC) under the European Union’s Horizon 2020 research and innovation programme (Grant agreement number 101001512).

\section*{Conflict of interest}
The authors declare no conflict of interest.

\section*{Keywords}
nanostructuring, direct laser writing, arbitrary optical potentials, optical potential tuning, thin films

\bibliographystyle{unsrt}
\bibliography{references}					

\begin{thebibliography}{10}

\bibitem{flatten2016}
L.C. Flatten, A.A.P. Trichet, and J.M. Smith.
\newblock Spectral engineering of coupled open-access microcavities.
\newblock {\em Laser \& Photonics Reviews}, 10(2):257--263, 2016.

\bibitem{Walker2021}
Benjamin~T. Walker, Benjamin~J. Ash, Aur\'{e}lien A.~P. Trichet, Jason~M.
  Smith, and Robert~A. Nyman.
\newblock Bespoke mirror fabrication for quantum simulation with light in
  open-access microcavities.
\newblock {\em Opt. Express}, 29(7):10800--10810, 2021.

\bibitem{trichet2015}
Aur\'{e}lien A.~P. Trichet, Philip~R. Dolan, David~M. Coles, Gareth~M. Hughes,
  and Jason~M. Smith.
\newblock Topographic control of open-access microcavities at the nanometer
  scale.
\newblock {\em Opt. Express}, 23(13):17205--17216, 2015.

\bibitem{greuter2014}
Lukas Greuter, Sebastian Starosielec, Daniel Najer, Arne Ludwig, Luc
  Duempelmann, Dominik Rohner, and Richard~J. Warburton.
\newblock A small mode volume tunable microcavity: Development and
  characterization.
\newblock {\em Applied Physics Letters}, 105(12):121105, 2014.

\bibitem{Epp2010}
E.~Epp, N.~Ponnampalam, W.~Newman, B.~Drobot, J.~N. McMullin, A.~F. Meldrum,
  and R.~G. DeCorby.
\newblock Hollow {B}ragg waveguides fabricated by controlled buckling of
  {Si/SiO2} multilayers.
\newblock {\em Opt. Express}, 18(24):24917--24925, 2010.

\bibitem{allen2011}
T.~W. Allen, J.~Silverstone, N.~Ponnampalam, T.~Olsen, A.~Meldrum, and R.~G.
  DeCorby.
\newblock High-finesse cavities fabricated by buckling self-assembly of
  {a-Si/SiO2} multilayers.
\newblock {\em Opt. Express}, 19(20):18903--18909, 2011.

\bibitem{Kurtscheid2020}
Christian Kurtscheid, David Dung, Andreas Redmann, Erik Busley, Jan Klaers,
  Frank Vewinger, Julian Schmitt, and Martin Weitz.
\newblock Realizing arbitrary trapping potentials for light via direct laser
  writing of mirror surface profiles.
\newblock {\em {EPL} (Europhysics Letters)}, 130(5):54001, 2020.

\bibitem{klaers2010}
Jan Klaers, Julian Schmitt, Frank Vewinger, and Martin Weitz.
\newblock {Bose--Einstein} condensation of photons in an optical microcavity.
\newblock {\em Nature}, 468(7323):545--548, 2010.

\bibitem{Schmitt2015}
Julian Schmitt, Tobias Damm, David Dung, Frank Vewinger, Jan Klaers, and Martin
  Weitz.
\newblock Thermalization kinetics of light: From laser dynamics to equilibrium
  condensation of photons.
\newblock {\em Phys. Rev. A}, 92:011602, 2015.

\bibitem{Vretenar2021_2}
Mario Vretenar, Chris Toebes, and Jan Klaers.
\newblock Modified {Bose-Einstein} condensation in an optical quantum gas.
\newblock {\em Nature Communications}, 12(1):5749, 2021.

\bibitem{kurtscheid2019}
Christian Kurtscheid, David Dung, Erik Busley, Frank Vewinger, Achim Rosch, and
  Martin Weitz.
\newblock Thermally condensing photons into a coherently split state of light.
\newblock {\em Science}, 366(6467):894--897, 2019.

\bibitem{Vretenar2021_1}
Mario Vretenar, Ben Kassenberg, Shivan Bissesar, Chris Toebes, and Jan Klaers.
\newblock Controllable {Josephson} junction for photon {Bose-Einstein}
  condensates.
\newblock {\em Physical Review Research}, 3(2):023167, 2021.

\bibitem{Toebes2022}
Chris Toebes, Mario Vretenar, and Jan Klaers.
\newblock Dispersive and dissipative coupling of photon {Bose-Einstein}
  condensates.
\newblock {\em Communications Physics}, 5(1):59, 2022.

\bibitem{busley2022}
Erik Busley, Leon~Espert Miranda, Andreas Redmann, Christian Kurtscheid,
  Kirankumar~Karkihalli Umesh, Frank Vewinger, Martin Weitz, and Julian
  Schmitt.
\newblock Compressibility and the equation of state of an optical quantum gas
  in a box.
\newblock {\em Science}, 375(6587):1403--1406, 2022.

\bibitem{scafirimuto2021}
Fabio Scafirimuto, Darius Urbonas, Michael~A. Becker, Ullrich Scherf, Rainer~F.
  Mahrt, and Thilo St{\"o}ferle.
\newblock Tunable exciton--polariton condensation in a two-dimensional {Lieb}
  lattice at room temperature.
\newblock {\em Communications Physics}, 4(1):39, 2021.

\bibitem{feng2022}
Feng Li, Yiming Li, L.~Giriunas, M.~Sich, D.~D. Solnyshkov, G.~Malpuech,
  A.~A.~P. Trichet, J.~M. Smith, E.~Clarke, M.~S. Skolnick, and D.~N.
  Krizhanovskii.
\newblock Condensation of {2D} exciton-polaritons in an open-access
  microcavity.
\newblock {\em Journal of Applied Physics}, 131(9):093101, 2022.

\bibitem{berden2000}
Giel Berden, Rudy Peeters, and Gerard Meijer.
\newblock Cavity ring-down spectroscopy: Experimental schemes and applications.
\newblock {\em International Reviews in Physical Chemistry}, 19(4):565--607,
  2000.

\bibitem{rotter2017}
Stefan Rotter and Sylvain Gigan.
\newblock Light fields in complex media: Mesoscopic scattering meets wave
  control.
\newblock {\em Rev. Mod. Phys.}, 89:015005, 2017.

\bibitem{Wlodarczyk2014}
Krystian~L. Wlodarczyk, Emma Bryce, Noah Schwartz, Mel Strachan, David Hutson,
  Robert R.~J. Maier, David Atkinson, Steven Beard, Tom Baillie, Phil
  Parr-Burman, Katherine Kirk, and Duncan~P. Hand.
\newblock Scalable stacked array piezoelectric deformable mirror for astronomy
  and laser processing applications.
\newblock {\em Review of Scientific Instruments}, 85(2):024502, 2014.

\bibitem{Canuel2012}
B~Canuel, R~Day, E~Genin, P~La Penna, and J~Marque.
\newblock Wavefront aberration compensation with a thermally deformable mirror.
\newblock {\em Classical and Quantum Gravity}, 29(8):085012, 2012.

\bibitem{bonora2012}
S.~Bonora, D.~Coburn, U.~Bortolozzo, C.~Dainty, and S.~Residori.
\newblock High resolution wavefront correction with photocontrolled deformable
  mirror.
\newblock {\em Opt. Express}, 20(5):5178--5188, 2012.

\bibitem{Dung2017}
David Dung, Christian Kurtscheid, Tobias Damm, Julian Schmitt, Frank Vewinger,
  Martin Weitz, and Jan Klaers.
\newblock Variable potentials for thermalized light and coupled condensates.
\newblock {\em Nature Photonics}, 11(9):565--569, 2017.

\bibitem{rubio1982}
F.~Rubio, J.~Denis, J.M. Albella, and J.M. Martinez-Duart.
\newblock Sputtered {Ta2O5} antireflection coatings for silicon solar cells.
\newblock {\em Thin Solid Films}, 90(4):405--408, 1982.

\bibitem{Cevro1995}
M.~Cevro.
\newblock Ion-beam sputtering of {(Ta2O5)x- (SiO2)1-x} composite thin films.
\newblock {\em Thin Solid Films}, 258(1):91--103, 1995.

\bibitem{chaneliere1998}
C.~Chaneliere, J.L. Autran, R.A.B. Devine, and B.~Balland.
\newblock Tantalum pentoxide {(Ta2O5)} thin films for advanced dielectric
  applications.
\newblock {\em Materials Science and Engineering: R: Reports}, 22(6):269--322,
  1998.

\bibitem{bartic2002}
Carmen Bartic, Henri Jansen, Andrew Campitelli, and Staf Borghs.
\newblock Ta2o5 as gate dielectric material for low-voltage organic thin-film
  transistors.
\newblock {\em Organic Electronics}, 3(2):65--72, 2002.

\bibitem{penn2003}
Steven~D Penn, Peter~H Sneddon, Helena Armandula, Joseph~C Betzwieser,
  Gianpietro Cagnoli, Jordan Camp, D~R~M Crooks, Martin~M Fejer, Andri~M
  Gretarsson, Gregory~M Harry, Jim Hough, Scott~E Kittelberger, Michael~J
  Mortonson, Roger Route, Sheila Rowan, and Christophoros~C Vassiliou.
\newblock Mechanical loss in tantala/silica dielectric mirror coatings.
\newblock {\em Classical and Quantum Gravity}, 20(13):2917, 2003.

\bibitem{zhao2003}
Yuanan Zhao, Yingjian Wang, Hui Gong, Jianda Shao, and Zhengxiu Fan.
\newblock Annealing effects on structure and laser-induced damage threshold of
  {Ta2O5/SiO2} dielectric mirrors.
\newblock {\em Applied Surface Science}, 210(3):353--358, 2003.

\bibitem{Paolone2022}
Annalisa Paolone, Ernesto Placidi, Elena Stellino, Maria~Grazia Betti, Ettore
  Majorana, Carlo Mariani, Alessandro Nucara, Oriele Palumbo, Paolo Postorino,
  Marco Sbroscia, Francesco Trequattrini, Massimo Granata, David Hofman,
  Christophe Michel, Laurent Pinard, Anaël Lemaitre, Nikita Shcheblanov,
  Gianpietro Cagnoli, and Fulvio Ricci.
\newblock Argon and other defects in amorphous {SiO2} coatings for
  gravitational-wave detectors.
\newblock {\em Coatings}, 12(7), 2022.

\bibitem{anghinolfi2013}
L~Anghinolfi, M~Prato, A~Chtanov, M~Gross, A~Chincarini, M~Neri, G~Gemme, and
  M~Canepa.
\newblock Optical properties of uniform, porous, amorphous {Ta2O5} coatings on
  silica: temperature effects.
\newblock {\em Journal of Physics D: Applied Physics}, 46(45):455301, 2013.

\bibitem{Paolone2021}
A.~Paolone, E.~Placidi, E.~Stellino, M.G. Betti, E.~Majorana, C.~Mariani,
  A.~Nucara, O.~Palumbo, P.~Postorino, I.~Rago, F.~Trequattrini, M.~Granata,
  J.~Teillon, D.~Hofman, C.~Michel, A.~Lemaitre, N.~Shcheblanov, G.~Cagnoli,
  and F.~Ricci.
\newblock Effects of the annealing of amorphous {Ta2O5} coatings produced by
  ion beam sputtering concerning the effusion of argon and the chemical
  composition.
\newblock {\em Journal of Non-Crystalline Solids}, 557:120651, 2021.

\bibitem{granata2020amorphous}
Massimo Granata, Alex Amato, Laurent Balzarini, Maurizio Canepa, J{\'e}r{\^o}me
  Degallaix, Dani{\`e}le Forest, Vincent Dolique, Lorenzo Mereni, Christophe
  Michel, Laurent Pinard, et~al.
\newblock Amorphous optical coatings of present gravitational-wave
  interferometers.
\newblock {\em Classical and Quantum Gravity}, 37(9):095004, 2020.

\bibitem{echlin2009}
Patrick Echlin.
\newblock {\em Handbook of Sample Preparation for Scanning Electron Microscopy
  and {X}-Ray Microanalysis}.
\newblock Springer, 2009.

\bibitem{Ivanda1991}
M.~Ivanda, K.~Furić, O.~Gamulin, M.~Peršin, and D.~Gracin.
\newblock {cw} laser crystallization of amorphous silicon: {Thermal} or
  athermal process.
\newblock {\em Journal of Applied Physics}, 70(8):4637--4639, 1991.

\bibitem{michaud2006}
J.F. Michaud, R.~Rogel, T.~Mohammed-Brahim, and M.~Sarret.
\newblock Cw argon laser crystallization of silicon films: Structural
  properties.
\newblock {\em Journal of Non-Crystalline Solids}, 352(9):998--1002, 2006.

\bibitem{Cinar2020}
Kamil {\c{C}}{\i}nar, Cihan Ye{\c{s}}il, and Alpan Bek.
\newblock Revealing laser crystallization mechanism of silicon thin films via
  pulsed {IR} lasers.
\newblock {\em The Journal of Physical Chemistry C}, 124(1):976--985, 2020.

\end{thebibliography}

\end{document}